\documentclass[twocolumn,showpacs,preprintnumbers,amsmath,amssymb,aps,prb]{revtex4}
\usepackage{graphicx}
\begin{document}

\title{
Colloidal Lattice Shearing and Rupturing with a Driven Line
of Particles} 
\author{
A. Lib{\' a}l$^{1,2}$, B.M. Cs{\' \i}ki$^{2}$, C. J. Olson Reichhardt$^{1}$, 
and C. Reichhardt$^{1}$ } 
\affiliation{
$^1$Theoretical Division,
Los Alamos National Laboratory, Los Alamos, New Mexico 87545 USA\\ 
$^2$ Faculty of Mathematics and Computer Science, Babes-Bolyai University, RO-400591 Cluj-Napoca, Romania 
} 

\date{\today}
\begin{abstract}
We examine the dynamics of two-dimensional colloidal systems 
using numerical simulations of a system
with a drive applied to a thin region in the middle of the sample to
produce a local shear.
For a monodisperse colloidal assembly, 
we find a well defined decoupling transition 
separating a regime of elastic motion from 
a plastic phase where the particles in the driven region 
break away or decouple from the particles in the bulk,
producing a shear band. 
For a bidisperse assembly, we find 
that the onset of a bulk disordering transition coincides  
with the broadening of the shear band. 
We identify several distinct dynamical regimes  
that are correlated with features in the velocity-force curves.
As a function of bidispersity, 
the decoupling force shows a nonmonotonic behavior associated
with features in the noise fluctuations, 
power spectra, and bulk velocity profiles.
When pinning is added in the bulk, we find 
that the shear band regions can become more localized, 
causing a decoupling of the driven particles from the bulk particles.   
For a system with thermal noise and no pinning,
the shear band region becomes 
more extended and the average 
velocity of the 
driven particles 
drops at the thermal disordering transition of the bulk system. 
\end{abstract}
\maketitle

\vskip2pc

\section{Introduction}
As the ability to manipulate small particles or groups of small
particles has advanced, there have been a growing number of studies examining 
the effect of local perturbations such as driving a single probe particle
through a particle assembly in
colloidal \cite{8,13,7,14,15,16,18,17,9,12,46,47,5,11} and  
granular \cite{45,10,3,4,6} media. 
The fluctuations and transport characteristics of the probe 
particle can undergo pronounced changes 
depending on the ordering of the
surrounding media, 
the particle-particle interactions, 
the temperature, and the magnitude of the
external force 
on the probe particle. 

In systems with short range interactions 
such as disks or granular particles \cite{45,6,10,3,4}, the velocity of
the probe particle can drop under even a small change in the
bulk media density
as the jamming transition is approached. 
For bulk densities below jamming,
the external force needed to move the probe particle is small or absent
and the probe particle velocity distribution is bimodal \cite{6}.
Close to jamming, the 
external force needed to push the 
probe particle through the medium becomes finite and rapidly increases, while
the velocity fluctuations are intermittent and have a power law
distribution that has been interpreted as arising due to criticality
associated with the jamming transition
\cite{4,45,6,10,13}. 

For systems with longer
range interactions such as charged
colloids, there 
is an elastic or coupled flow regime at low drives where
the probe particle drags the entire assembly of particles
without any local tearing 
\cite{1}. 
At higher drives there can be a well defined
transition to a plastic response 
regime where the probe particle induces tearing
rearrangements in the surrounding media, and the velocity of the
probe particle differs from that of the particles it is dragging.
The velocity-force curves in these systems
can be 
used to identify the
critical driving force 
separating the elastic 
and plastic regimes.
The velocity increases as a power law with increasing external force
in the plastic   
regime, 
similar to the behavior 
observed for plastic depinning of vortices \cite{43} 
and colloids \cite{50} in the presence of quenched disorder.

It is also possible to consider the motion of a forced particle through
a surrounding medium in the presence of quenched disorder.
Such a situation has been 
studied for dragging individual vortices 
in type-II superconductors \cite{24,25,26,27}, 
where a driven vortex interacts with the intrinsic pinning 
in the sample as well as with
the surrounding vortices.
In this case, the probe particle can be 
indirectly pinned due to its interaction with bulk particles that are
strongly directly pinned by the substrate, causing the appearance of a
threshold depinning force for the probe particle \cite{16}.
At higher drives, the bulk pinning can counterintuitively
lower the effective drag on the probe particle.  This occurs since the drag
originates when having a portion of the surrounding particles move with the
probe particle, so that when the quenched disorder strongly immobilizes
the surrounding particles, they can no longer move with the probe particle,
leading to an effective decoupling of the 
probe particle from the surrounding media \cite{16}.
For probe particles driven through crystalline structures in the 
absence of quenched disorder, 
shear thinning \cite{8} and directional locking effects 
occur where the amount of drag on the
probe particle depends on the orientation of 
the drive with respect to the symmetry directions of
the crystalline structure. 
Flow along a symmetry direction
produces a lower drag since the probe particle can 
move easily between particles in the
background lattice without constantly generating new topological defects, 
while for other driving directions, the probe particle
induces the formation of
local topological defects, leading to increased drag \cite{49}.  
It was also shown that
for driving along symmetry directions, 
the drag increases as the bulk melting temperature
is approached due to the formation of topological defects in
the region surrounding the probe particle, leading to a local melting
transition \cite{14}.  
      
Here, instead of a single driven probe particle,
we consider the case of a localized 
quasi-one-dimensional region of particles driven externally along a line.
The drive is parallel to the orientation of the line.
Such a geometry could be created using a laser beam 
directed along the sample edge, with
magnetic particles driven by  a magnetic strip, 
or with particles coupled to a mechanical external drive
or driven with one-dimensional arrays of optical traps \cite{29,30,31}. 
Driven line geometries have already been 
realized for particles with Yukawa interactions in dusty plasmas, where
a laser beam pushes particles only along a line \cite{19,20,21,22,23}. 
It is also possible to create systems with 
bulk pinning and an easy flow channel in confined geometries 
\cite{28,34,35,40}. 
A key difference between the
point probe particle and the line probe 
considered here is that the line probe more directly 
mimics a shear response. If the system is strongly coupled, 
then an elastic response will occur in which the particles in the system
move along with the particles in the driven line. 
Here we study the case of 
a line of particles driven along a symmetry direction 
of the background lattice for 
monodisperse ordered, bidisperse ordered, and 
bidisperse disordered systems,
both for pin-free samples and for samples that
have pinning present in the bulk.      

\begin{figure}
\includegraphics[width=3.5in]{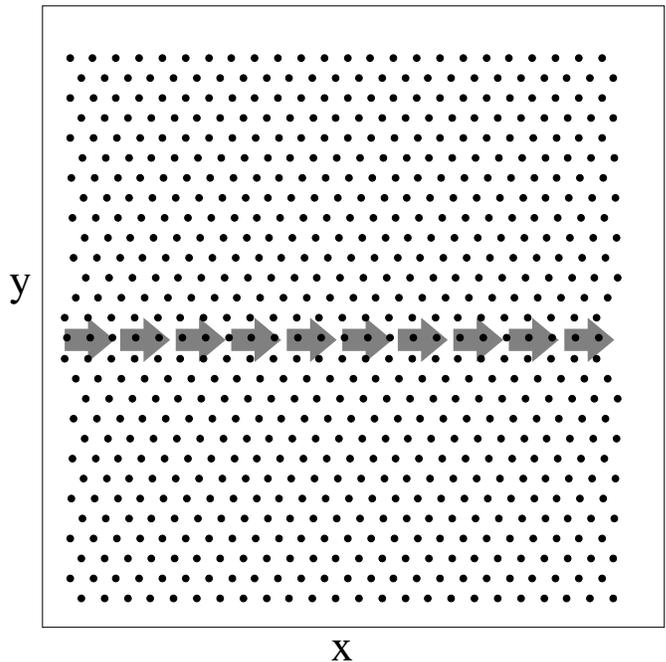}
\caption{
Illustration of the sample geometry for the monodisperse case.
Black dots:  Colloids interacting via a pairwise Yukawa potential to form
a triangular lattice.
Arrows indicate the region within which an external drive in the positive $x$
direction is applied to the particles.
There are periodic boundary conditions in the $x$ and $y$-directions.  
}
\label{fig:1}
\end{figure}

\section{Simulation and System}   
In Fig.~\ref{fig:1} we illustrate the geometry of the sample,
which 
consists of a two-dimensional assembly of 
$N_{c}$ colloids with periodic boundary conditions 
in the $x$ and $y$-directions. 
The arrows in the middle of Fig.~\ref{fig:1} indicate 
the region within which 
the colloids experience
an external drive ${\bf F}^D=F^D{\bf \hat x}$.
Colloids outside this region are undriven and have $F^D=0$.
The colloid density is $\rho = N_{c}/L^2$, 
where $L$ is the size of the simulation cell.
The particles interact 
via a Yukawa or screened Coulomb potential. 
In some cases, we also consider the
effect of pinning sites that we model as localized 
parabolic traps of radius $r_{p}$.

The  dynamics of particle $i$ arise from integrating 
the following equation of motion:  
\begin{equation}  
\eta \frac{d {\bf R}_{i}}{dt} = 
-\sum_{i\ne j}^{N_{i}}{\bf \nabla}V(R_{ij}) +  {\bf F}^{P}_{i} +  {\bf F}^{D}_{i} +  {\bf F}^{T}_{i}. 
\end{equation} 
Here $\eta$ is the damping constant, 
${\bf R}_{i(j)}$ is the position of particle $i(j)$,
and $R_{ij} = |{\bf R}_{i} - {\bf R}_{j}|$. 
The particle-particle interaction potential is
$V(R_{ij}) =  q^2E_{0}\exp(-\kappa R_{ij})/R_{ij}$.   
Here $E_{0} = Z^{*2}/4\pi\epsilon\epsilon_{0}a_{0}$, where 
$q$ is the dimensionless interaction strength,
$Z^{*}$ is the 
effective charge of the colloid,
$\epsilon$ is the solvent dielectric constant, 
and $1/\kappa$ is the screening length.  
Lengths are measured in units of $a_{0}$, time in units of 
$\tau = \eta/E_{0}$, and forces in units of $F_{0} = E_{0}/a_{0}$.
The driving force ${\bf F}^{D}=F_d{\bf \hat x}$ is applied only to particles 
in a region of width $d$ at the center of the sample.
We increase
the external drive in increments of $\delta F_{D}$, 
then hold the drive at a constant value 
for a fixed period of time and measure the average velocity
of the particles within the driven channel $V_{c}$ 
and of the particles in the bulk $V_{b}$.
The pinning force arises from $N_{p}$ non-overlapping parabolic pinning 
sites that are placed outside the driven channel.
The pinning force has the form 
${\bf F}^{P}_{i} = F_{p}(R_{ik}/R_{p})\Theta(R_{p}- R_{ik}){\bf \hat{R}}_{ik}$, 
where $R_{ik} = |{\bf R}_{i} - {\bf R}_{k}|$ is the distance between particle 
$i$ and the center of pinning site $k$, and
${\bf {\hat R}}_{ik} = ({\bf R}_{i} - {\bf R}_{k})/R_{ik}$. 
Here $R_{p}$ is the pinning site radius, $F_{p}$ is the maximum force of the 
pinning site, and $\Theta$ is the Heaviside step function.   
The effects of thermal fluctuations come from the 
Langevin noise term $F^{T}$ with the properties
$\langle F^{T}(t)\rangle = 0$ and 
$\langle F^{T}_{i}(t)F_j^T(t^{\prime})\rangle = 2\eta k_{B}\delta_{ij}\delta(t - t^{\prime})$,   
where $k_{B}$ is the Boltzmann constant.
The initial particle configurations are obtained by placing the colloids 
in a triangular arrangement, and unless otherwise noted, the average
lattice constant is $a=2.0$. 

\begin{figure}
\includegraphics[width=3.5in]{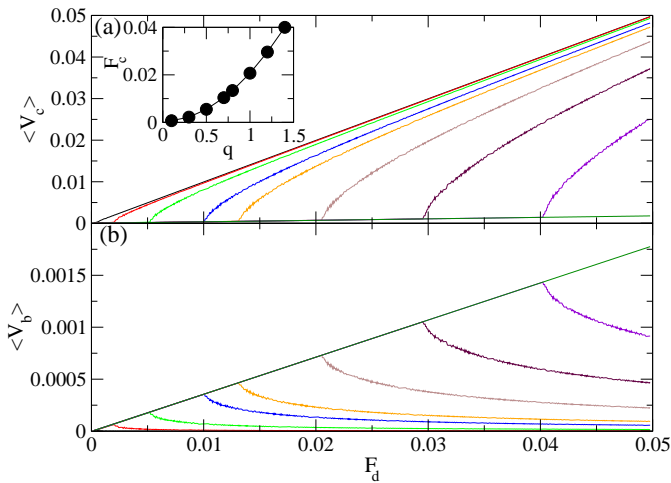}
\caption{
A monodisperse system of colloids 
with
varied interaction strength $q$ and with $a=2$.
(a) The velocities of the particles in the driven channel region 
$\langle V_{c}\rangle$ vs $F_{d}$, where the velocities 
are normalized by the number of particles in the driven region.
(b) The velocities of the particles in the bulk outside of the
driven region $\langle V_{b}\rangle$ vs $F_{d}$, 
where the velocities are normalized by the number of particles 
in the non-driven region.
The curves are for $q$ values (from left to right)  $q=0.1$, 
0.3, 0.5, 0.7, 0.8, 
1.0, 1.2, 1.4, and $1.6$. 
There are two regimes.  The first is an elastic region with $V_c=V_b$ where
all the particles move together.
This is followed at higher $F_d$
by a decoupled regime where the particles in the 
driven line move past the bulk particles, while the bulk particles
remain locked with each other.
The inset in (a) shows that the driving force $F_{c}$ at 
which the decoupling transition occurs increases with increasing $q$. 
}
\label{fig:2}
\end{figure}

\section{Monodisperse System} 

In Fig.~\ref{fig:2}(a) we plot the average velocity of the particles in 
the driven channel region $\langle V_c\rangle$ versus 
external drive $F_{d}$, and in Fig.~\ref{fig:2}(b) we plot
the corresponding average velocity of the particles outside 
of the driven channel, $\langle V_{b}\rangle$, versus $F_{d}$ 
for a system with a monodisperse triangular crystalline arrangement
of colloids. The different curves indicate the effect of changing
the interaction coefficient $q$. 
For low $F_d$, all the curves for both $\langle V_c\rangle$ and
$\langle V_b\rangle$ increase linearly and 
$\langle V_c\rangle=\langle V_b\rangle$,
corresponding to the elastic flow regime where the particles 
in the driven region are locked 
with the bulk particles. 
Here there is no tearing or rearrangements, 
so all the particles keep 
their same neighbors and move together. 
At higher drives, a transition occurs to a regime where $\langle V_b\rangle$
decreases and $\langle V_c\rangle$ increases with increasing $F_d$.
This occurs
when the particles in the driven channel partially decouple from
the bulk particles and are able to move past them. 
The bulk particles continue to move due to the coupling to the driven
particles; however, as $F_d$ increases, this coupling becomes weaker and
$\langle V_b\rangle$ decreases.
The
inset in Fig.~\ref{fig:2}(a) shows the force $F_{c}$ at which 
the decoupling occurs as a function of $q$.
As the strength of the particle-particle interactions increases, a larger
force must be applied before decoupling can occur.

The velocity-force curves in Fig.~\ref{fig:2} 
are very similar to those found in other
systems that exhibit decoupling and drag effects.
One of the best known examples is the Giaever transformer geometry 
for coupled superconducting layers in the presence of a 
magnetic field \cite{38,39}.  Each quantized magnetic field line must
pass through both
layers, providing a magnetic coupling between the layers.
When only one layer is driven by a current, the voltage response, which is 
proportional to the 
superconducting vortex velocity, is identical in both layers,
indicating that the vortices in the two layers are locked together. 
As the drive increases, there
is a decoupling transition between the layers associated with a decrease
in the effective drag on the driven layer.  This causes the velocity of
the vortices in the driven layer to increase while the velocity of the
vortices in the secondary or undriven layer drops.
In this superconducting system, the interaction between vortices in
neighboring layers is attractive, while
in the colloidal system, the interactions between colloids in the driven and
undriven regions are repulsive. 
The response of repulsively interacting particles in coupled one-dimensional
wires has been studied for
classical electrons \cite{51} and 
particles with Yukawa interactions \cite{41,44}. In these systems, 
a commensurate state can form when the
number of particles in each layer is the same,
producing
a well defined coupling-decoupling transition when one layer is driven. 
In the case of our driven line of particles,
the system can be viewed as a single driven 
one-dimensional layer interacting with an array of
non-driven layers.
At the decoupling transition, the particles 
outside of the driven region remain locked together so 
that a very small shear band occurs only at the driven line. 
For more disordered systems,
a much broader shear band forms.    

\begin{figure}
\includegraphics[width=3.5in]{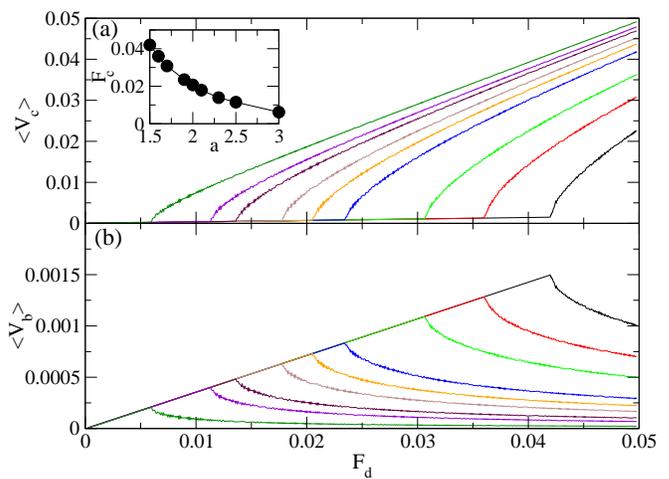}
\caption{ 
A monodisperse system of colloids with $q=1.0$ for varied colloid density.
(a) $\langle V_{c}\rangle$ vs $F_{d}$. (b) 
$\langle V_{b}\rangle$ vs $F_{d}$. 
The colloid lattice constant $a =  3.0$, 2.5, 2.3, 2.1, 2.0, 1.9, 
1.7, 1.6, and $1.5$, from left to right. 
Samples with larger $a$ are less dense and decouple at a lower driving
force.
The inset in (a) shows 
$F_c$ vs $a$.
}
\label{fig:3}
\end{figure}
 
In Fig.~\ref{fig:3} we show $\langle V_{c}\rangle$ and 
$\langle V_{b}\rangle$ vs $F_{d}$ for a 
monodisperse sample with fixed $q = 1.0$ but varied colloid density, 
measured in terms of the lattice constant $a$.
The inset of Fig.~\ref{fig:3}(a) shows the dependence of the decoupling force $F_c$
on $a$. 
We observe a trend similar to that of changing $q$, where the samples with
lower density have weaker particle-particle interactions and a lower
decoupling threshold.
The decoupling transition should also 
depend on
the commensurability ratio of the particles in the driven channel 
with the surrounding media. If the density in the driven channel is
higher or lower than that of the bulk, the decoupling force will be depressed
relative to the case we consider here where the two densities are equal.
This is a result of the localized incommensurations that separate the two
regions of different densities.  The incommensurations depin below the
bulk decoupling transition and shift the transition to lower $F_d$.
Studies of one-dimensional coupled channels
also found that the decoupling transition drops to lower drives
at incommensurate channel filling ratios \cite{41}.     
 
\section{Bidisperse Systems}

\begin{figure}
\includegraphics[width=3.5in]{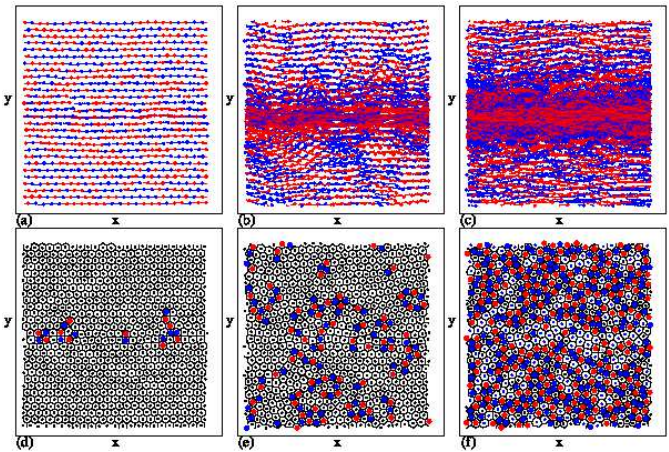}
\caption{
A bidisperse colloidal system 
where half the particles 
have charge $q_{1}$ and the other half have charge $q_{2}=1.0$. 
(a,b,c) Dots: Particle positions.  Dark (blue) particles have charge
$q_1$ and light (red) particles have charge $q_2$.  Lines indicate the 
trajectories followed by the particles over a fixed time interval.
(d,e,f) Voronoi construction. 
White polygons correspond to particles with six neighbors, dark blue polygons
to 
particles with five neighbors, and 
light red polygons to particles with seven neighbors. 
(a,d) At $q_1/q_{2} = 0.7$ and $F_d=0.05$, the system is
mostly ordered.  There are some dislocations 
near the driven region due to the fact that the driven particles are moving
faster than the bulk particles.  These dislocations all 
have their Burgers vectors
aligned in the same direction.
(b,e) At $q_{1}/q_{2} = 1.6$, the system is partially disordered in the bulk 
and there
is a mixing of the trajectories as the shear band widens. 
(c,f) At $q_{1}/q_{2} = 2.4$, the system is strongly disordered
and the width of the shear band is increased.
}
\label{fig:4}
\end{figure}

We next consider a bidisperse system where half the 
particles have an interaction coefficient of $q_{1}$
and the other half have an interaction coefficient of $q_{2}$. 
In Fig.~\ref{fig:4}(a,b,c) we plot the particle positions and trajectories during
a fixed period of time, and in   
Fig.~\ref{fig:4}(d,e,f) we show the Voronoi tessellations of the particle positions 
at one instant.
For $q_{1}/q_{2} = 0.7$ 
and $F_{d} = 0.05$, in Fig.~\ref{fig:4}(a,d), 
the system is crystalline with sixfold ordering
in the bulk. The topological defects are concentrated near the 
driven line since the system is in the
decoupled phase where the particles in the 
driven line move faster than the bulk particles. 
There is also 
no transverse diffusion in the system, as 
indicated by the nearly one-dimensional, non-mixing 
trajectories of the particles.        
In Fig.~\ref{fig:4}(b,e) at $q_{1}/q_{2} = 1.6$, the bulk is disordered
with a proliferation of 5-7 paired defects. 
The trajectories show that there is strong mixing near
the driven line, with the most strongly disordered region 
concentrated near the driven line. 
Fig.~\ref{fig:4}(c,f) shows that for $q_{1}/q_{2} = 2.4$, 
the system is even more disordered, and contains an
increased number of 5-7 defects.
In addition, the region of strong mixing, denoted by the region with
crossing trajectories, is now wider in the $y$-direction.
For $0.6 < q_1/q_2 < 1.6$, the bulk is ordered.
For $q_{1}/q_{2} < 0.6$ the system disorders 
and the trajectories are similar to those shown in Fig.~\ref{fig:4}(b).

\begin{figure}
\includegraphics[width=3.5in]{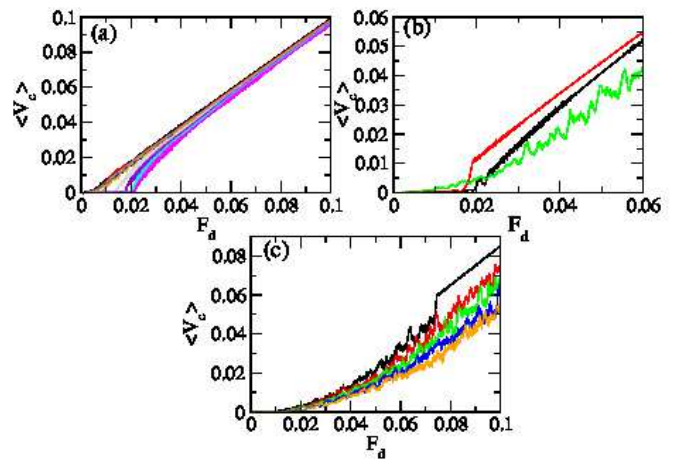}
\caption{ 
$\langle V_{c}\rangle$ vs $F_{d}$ for bidisperse systems with $q_2=1.0$.
(a) $q_1/q_{2} = 0.1$, 0.2, 0.3, 0.5, 0.7, 0.8, 0.9, 1.0, and $1.1$, from 
left to right.
The system is disordered for $q_1/q_{2} \leq 0.6$. 
(b) $q_1/q_{2} = 1.3$ (center right), 1.4 (upper right), and $1.8$
(lower right). $F_{c}$ is highest for $q_1/q_{2} = 1.3$ and
lowest for $q_1/q_{2} = 1.8$. 
The system is disordered for 
$q_1/q_{2} = 1.8$, producing the enhanced fluctuations in the 
velocities. 
There is also a crossing of the velocity-force curves,
where $\langle V_c\rangle$ 
at $q_1/q_{2} = 1.8$ is lower in the moving phase than 
$\langle V_c\rangle$ of the ordered state due to the widened shear band in the
moving state in the disordered system.
(c) The disordered states at $q_1/q_{2} = 2.2$, 2.4, 2.6, 2.8 and $3.0$, 
from top to bottom.  Here $\langle V_c\rangle$ for fixed $F_d$ decreases
with increasing $q_1/q_2$.
}
\label{fig:5}
\end{figure}

The different phases can also be 
identified via changes in the features of the velocity-force curves. 
In Fig.~\ref{fig:5}(a) we 
plot $\langle V_{c}\rangle$ vs 
$F_{D}$ for samples with $q_{1}/q_{2} = 0.1$ to  $1.1$.
As $q_{1}/q_{2}$ increases, the decoupling force $F_{c}$ increases. 
The concavity of the
velocity-force curves can be fit to $\langle V_c\rangle \propto (F-F_c)^\beta$,
where $\beta > 1.0$ for the disordered systems with  
$q_1/q_2\leq 0.6$ 
and
$\beta<1.0$ in the ordered systems.
This is shown more clearly in
Fig.~\ref{fig:5}(b) where we plot $\langle V_{c}\rangle$ vs $F_{d}$ for 
systems with $q_{1}/q_{2} = 1.3$, 1.4, and $1.8$.
For the disordered case of $q_1/q_2=1.8$, there are stronger fluctuations
in $\langle V_c\rangle$.
$F_{c}$ is highest for $q_{1}/q_{2} = 1.3$ but 
is lower in the disordered regime for $q_{1}/q_{2} = 1.8$. 
A crossing of the velocity-force curves occurs since $F_c$ for the
$q_1/q_2=1.8$ system is lower than for the $q_1/q_2=1.3$ system,
but at higher drives $\langle V_c\rangle$ is lower in the $q_1/q_2=1.8$
sample.
This occurs since the ordered state at
$q_{1}/q_{2} = 1.3$ produces a sharp shear band,
while the disordered state at $q_{1}/q_{2} = 1.8$ has a wide shear band
region similar to that illustrated in Fig.~\ref{fig:4}(b), 
meaning that more particles are being dragged by the line particles,
causing the driven particles to move more slowly. 

\begin{figure}
\includegraphics[width=3.5in]{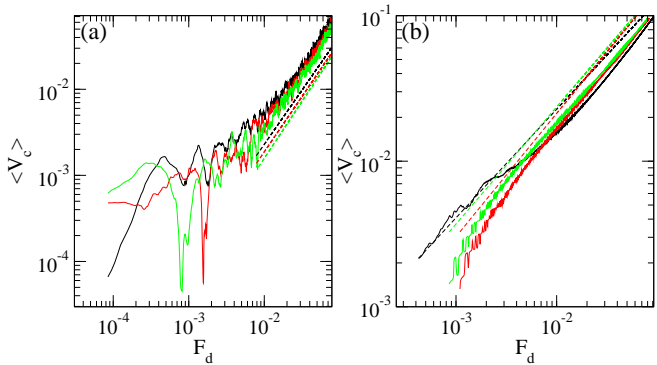}
\caption{ 
(a) $\langle V_{b}\rangle$ vs 
$F_{d} - F_{c}$ for a bidisperse system with $q_2=1.0$,
where $F_{c}$ is the force at decoupling or depinning. 
(a) $q_{1}/q_{2} = 2.6$, 2.8, and $3.0$, from top right to bottom right. 
The dashed lines are power law fits
(vertically offset for clarity) with exponents of $\beta = 1.26$, 1.29, and 
$1.32$, 
respectively. 
(b) The same for the ordered regime of 
$q_{1} = 0.7$, $0.9$, and $1.0$, 
from bottom right to top right, with dashed lines indicating power law fits 
(vertically offset for clarity)
of $\beta = 0.746$, 0.84, and $0.8$, respectively.  
}
\label{fig:6}
\end{figure}

Figure~\ref{fig:5}(c) shows $\langle V_{c}\rangle$ versus $F_{D}$ 
for $2.2 \leq q_{1}/q_{2} \leq 3.0$, where
the system remains disordered with large fluctuations in 
$\langle V_c\rangle$. 
For a fixed drive, $\langle V_c\rangle$
drops with increasing $q_{1}/q_{2}$. 
Here the velocity-force curves can
be fit to the form 
$\langle V_{c}\rangle - V_{dc} \propto (F_{d} - F_{c})^\beta$ 
with $\beta \approx 1.3$, where
$V_{dc}$ is the velocity at $F_{c}$. This fit is
illustrated in Fig.~\ref{fig:6}(a) for $q_{1}/q_{2} = 2.6$, 2.8, and $3.0$. 
A power law fit of this type
is similar to that observed for a single particle 
moving through a disordered medium when the
driven particle creates a large amount of distortion, including
particle rearrangements \cite{1,13}.
Similar scaling is found at depinning for assemblies of particles that
undergo plastic flow upon depinning; such tearing behavior generates
an exponent ranging from $\beta=1.25$ to $\beta=2.0$ \cite{43,50}.
Near an elastic depinning transition, the velocity-force curve can
also be fit with a power law but with a smaller exponent 
$\beta < 1.0$ \cite{42,50}. 
For samples in the ordered regime, we can fit $\langle V_c\rangle - V_{dc}$
to a similar scaling form with
$\beta \approx 0.8$, as shown in Fig.~\ref{fig:6}(b) for 
$q_1/q_2 = 0.7$, 0.9, and $1.0$.
This indicates that the decoupling transition in the ordered phase 
resembles the elastic depinning
of a one-dimensional coupled chain of particles 
from a periodic substrate created by the
periodic ordering of the bulk particles. 
We find similar scalings for the other fillings we have considered, 
where for the ordered systems $\beta < 1.0$ and 
for the disordered systems $\beta > 1.0$.

\begin{figure}
\includegraphics[width=3.5in]{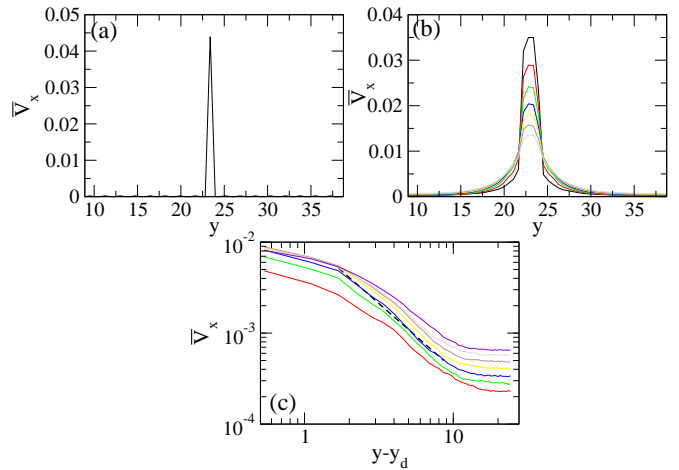}
\caption{ 
(a,b) Velocity profiles $\bar V_x$ vs $y$ for a
bidisperse system with $q_2=1.0$.
The driven region is centered at $y_d=23$.
(a) 
In a sample with $q_1/q_2=1.0$ at $F_d=1.0$,
a sharp shear band forms.
(b) Samples with
$q_{1}/q_{2} = 1.8$, 2.0, 2.2, 2.4, 2.6, 2.8, and $3.0$, 
from top center to bottom center. 
$\bar V_x$ in the driven region drops with increasing $q_1/q_2$, while
$\bar V_x$ in the bulk increases 
with increasing $q_1/q_2$ as the shear band widens.
(c) $\bar V_x$ in only the top half of the sample vs $y-y_d$ for
samples with
$q_{1}/q_{2} = 1.8$, 2.0, 2.2, 2.4, 2.6, 2.8, and $3.0$, from top to bottom.
Dotted line is a power law fit for the $q_1/q_2=2.2$ system with
an exponent of
$-1.4$ 
in the region of strong mixing.
At large $y-y_d$, the bulk becomes locked again and moves as a solid.       
}
\label{fig:7}
\end{figure}

The shear banding effect can be better seen by examining profiles of the
average $x$ velocity $\bar V_x$ 
taken at different points along the $y$-direction.
Here, $\bar V_x(y) = \sum_{i=1}^{N_c} ({\bf \dot R}_i \cdot {\bf \hat x})[\Theta(R_y^i-y-\delta y/2)
\Theta(y+\delta y/2 -R_y^i)]$ where 
$R_y^i$ is the $y$ coordinate of particle $i$ and $\delta y$ is the
width of the averaging region.
Figure~\ref{fig:7}(a) shows the velocity profiles $\bar V_x(y)$
for $q_{1}/q_{2} = 1.0$ at 
$F_{d}  = 1.0$. A sharp spike in $\bar V_x$ appears at $y=y_d$, the center
of the driven region, while $\bar V_x \approx 0$ in the bulk undriven
portion of the sample.
This type of profile is observed for all samples where the 
system remains ordered in the decoupled phase.
In Fig.~\ref{fig:7}(b) we plot $\bar V_x(y)$
for disordered samples with
$q_{1}/q_{2} = 1.8$, 2.0, 2.2, 2.4, 2.6, 2.8, and $3.0$.
In all cases, $\bar V_x$ is maximum at $y=y_d$ and falls off for larger
$|y-y_d|$.
As $q_1/q_2$ increases,
$\bar V_x$ at $y=y_d$ decreases while in the bulk
$\bar V_x$ increases.
This coincides with the 
decrease in $\langle V_{c}\rangle$ for increasing $q_{1}/q_{2}$
shown in Fig.~\ref{fig:5}(c). 
The increase of the response in the bulk occurs when a larger number
of particles are dragged by the driven line
due to the increased particle-particle interaction strength.
In Fig.~\ref{fig:7}(c) we plot $\bar V_x$ for only the upper half of the sample
versus $y-y_d$ on a log-log scale, to better show where the shear banding
is occurring.
The dashed line in Fig.~\ref{fig:7}(c) is a power law fit to the data from the
$q_1/q_2=2.2$ sample in the shear band region, and has an exponent of
$-1.4$.
From the trajectory images in Fig.~\ref{fig:4}(c), it is clear that there is
a region of strong mixing near the driven line and a region of less mixing
at larger values of $y-y_d$, correlated with
a saturation of $\bar V_x$ in Fig.~\ref{fig:7}(c).
As $q_{1}/q_{2}$ increases, the saturation region 
shifts out to larger values of $y-y_d$. 
This results shows that the system has
liquid-like behavior near the shear band region, 
but acts more like a disordered elastic solid away from the driven region.

\begin{figure}
\includegraphics[width=3.5in]{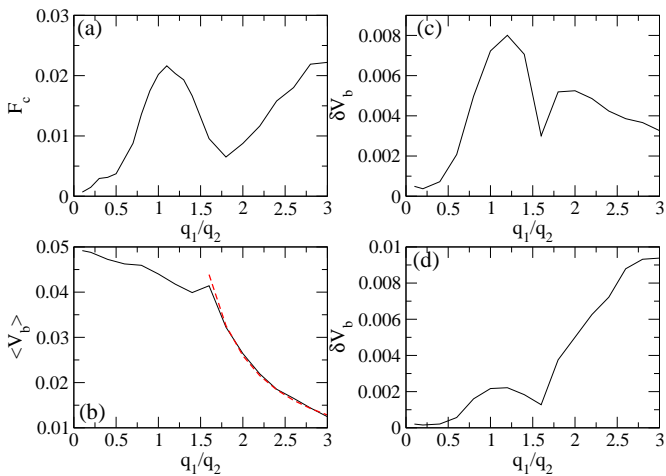}
\caption{ 
A bidisperse sample with $q_2=1.0$.
(a) $F_{c}$ vs $q_{1}/q_{2}$.
A peak appears in the ordered region
centered close to $q_{1}/q_{2} = 1.0$. (b) 
$\langle V_{b}\rangle$ vs $q_{1}/q_{2}$ 
at $F_{d} = 0.05$.
The velocity drops 
in the disordered regime
$q_1/q_2\geq 1.6$.  The dashed line indicates
a fit to
$1/(q_{1}/q_{2})$ in the disordered regime.  
(c) The standard deviation of the 
fluctuations in the bulk velocity $\delta V_{b}$ 
vs $q_{1}/q_{2}$ for $F_{d} =0.05$.
(d) $\delta V_{b}$ vs $q_{1}/q_{2}$ for $F_{d} = 0.1$. 
}
\label{fig:8}
\end{figure}

We summarize the behavior of 
$F_{c}$ vs $q_{1}/q_{2}$
for the bidisperse samples 
in Fig.~\ref{fig:8}(a). 
In the disordered regime 
$0.1 < q_{1}/q_{2} \leq 0.6$, 
$F_{c}$ increases with increasing $q_1/q_2$. 
Above $q_1/q_2=0.6$, when the system enters the ordered 
regime, $F_c$ increases more rapidly 
and the crystalline order that appears near $q_1/q_2=1.0$
enhances the coupling transition, 
as indicated by the peak in $F_{c}$ near $q_{1}/q_{2} = 1.0$. 
As the system becomes disordered away from $q_1/q_2=1.0$,
weak or defected spots appear that lower the decoupling transition
and reduce $F_c$.
A minimum in $F_c$ appears near $q_{1}/q_{2} = 1.8$ 
in the disordered region, while
$F_{c}$ increases again for larger $q_{1}/q_{2}$ 
due to the increasing strength of the particle-particle
interactions, which also cause the system to depin plastically.
The inset of Fig.~\ref{fig:2}(a) shows
that $F_{c}$ in the monodisperse system also increases with increasing $q$. 

In Fig.~\ref{fig:8}(b) we plot $\langle V_{b}\rangle$ 
vs $q_{1}/q_{2}$ in the bidisperse samples at $F_{d} = 0.05$. 
Here the bulk velocity decreases
at a moderate rate with increasing $q_1/q_2$
until the system becomes disordered 
for $q_1/q_2 \geq 1.6$ and 
large shear bands form, leading to a more rapid drop
in $\langle V_{b}\rangle$ with increasing $q_1/q_2$. 
The dashed line is a fit to $1/(q_1/q_{2})$
in the disordered regime. 
Just before the system disorders, 
for $1.4 \leq q_1/q_2 < 1.6$, there is 
a slight increase in $\langle V_b\rangle$ that is
correlated with a drop in the standard deviation of the
velocity fluctuations $\delta V_b$, as shown in Fig.~\ref{fig:8}(c).
For $F_{d} = 0.05$, Fig.~\ref{fig:8}(c) indicates that $\delta V_b$ reaches
a maximum for $q_1/q_2=1.1$, the same value at which there is a
peak in $F_c$ in Fig.~\ref{fig:8}(a).
At this point, when the velocity oscillations 
are the largest, the particles are moving in a one-dimensional
ordered pattern and each particle exhibits an oscillating velocity component
due to the effective periodic potential created by the ordered particles
in the bulk.
There is a dip in $\delta V_b$ at the transition from the ordered
to the disordered region at $q_1/q_2=1.6$; for $q_1/q_2>1.6$, 
$\delta V_b$ decreases with increasing $q_1/q_2$.
At the disordering transition, the 
particles no longer move together but have different 
sliding velocities, resulting in a cancellation of the large oscillations
that appeared in the ordered phase.
For larger $q_1/q_2$, deeper into the disordered regime,
the velocity fluctuations decrease with increasing $q_1/q_2$ due to a
decrease in the amount of motion in the system as the fixed value of $F_d$ 
gets closer to $F_c$, which increases with increasing $q_1/q_2$. 
At $F_{d} = 0.1$, Fig.~\ref{fig:8}(d) shows that the same general
trends in $\delta V_b$ persist;
however, in this case the 
velocity fluctuations for larger $q_1/q_2$ increase with
increasing $q_{1}$ since the higher drive causes 
a large shear banding effect. 
There is a saturation of $\delta V_b$ near
$q_{1}/q_{2} = 3.0$, and we expect that the velocity fluctuations will
decrease for higher $q_1/q_2$ as $F_c$ increases and approaches the
fixed value $F_d=0.1$.
These results indicate that the order-disorder transition can be detected
by measuring the velocity fluctuations.

\begin{figure}
\includegraphics[width=3.5in]{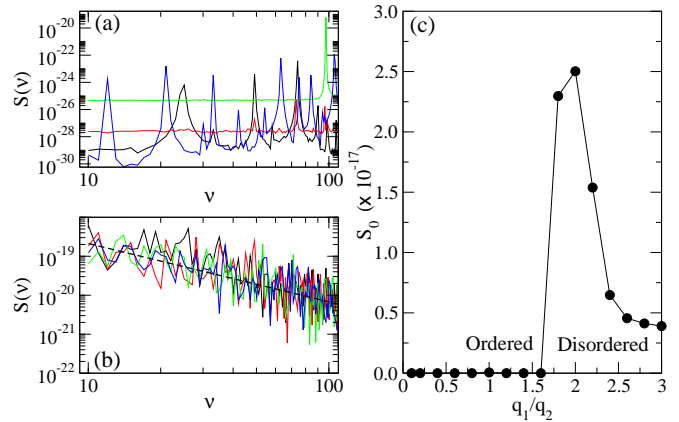}
\caption{ 
(a,b) Power spectra $S(\nu)$ for  bidisperse samples with
$q_2=1.0$ 
at $F_d=0.05$.
(a) In ordered systems, 
$q_{1}/q_{2} = 0.6$ (black), 0.8 (red), 1.0 (green), and $1.4$ (blue), 
there is a narrow band noise signature.
(b) In disordered systems, 
$q_{1}/q_{2} = 2.4$ (black), 2.6 (red), 2.8 (green), and $3.0$ (blue), 
a broad band noise signal occurs.
The dashed line
is a power law fit with an exponent of $-1.5$.
(c) The noise power $S_0$ vs $q_1/q_2$.  At the transition into the
disordered regime, $S_0$ jumps to a much higher value.
}
\label{fig:9}
\end{figure}

The velocity fluctuations can also be characterized using the
power spectrum 
$S(\nu) = |\int V_b(t)e^{-2\pi i\nu t}dt|^2$. 
In the ordered regime, the noise fluctuations 
have a narrow band noise feature indicating 
that there is a characteristic frequency. 
Similar narrow band noise occurs for particles moving over a 
periodic substrate, where the characteristic frequency is determined by
the rate at which the particle travels from one substrate minimum to
the next \cite{52}.
In Fig.~\ref{fig:9}(a) we plot $S(\nu)$ in the ordered regime for
samples with 
$q_{1}/q_{2} = 0.6$, 0.8, 1.0, and $1.4$ for
a fixed drive of $F_{d} = 0.05$. 
The peaks in the spectra indicate the presence of narrow band noise.
In the disordered regime, we find broad band noise or a
$1/f^{\alpha}$ noise signature, as shown in Fig.~\ref{fig:9}(b) for
$q_1/q_2=2.4$, 2.6, 2.8, and 3.0.  Here $\alpha =-1.5$.  Similar broad
band noise has been observed
for dragging a single particle through granular media at the 
jamming transition \cite{16} and for plastic depinning of particles
on disordered substrates \cite{53}.    
We can also analyze the noise power 
$S_{0}=\int_{\nu_1}^{\nu_2}S(\nu)$ for a fixed 
octave, as shown in Fig.~9(c) 
for $\nu_1=10$ and $\nu_2=100$.
Here $S_0$ is small in the ordered regime $0.6 < q_1/q_2 < 1.6$ and undergoes
a pronounced increase to a maximum near the onset of the disordered
regime.
For $q_{1}/q_{2} \leq 0.6$, 
the system is also disordered and $S(\nu)$ shows broad band noise; however,
the noise power remains very low for these small values of $q_1/q_2$.

\begin{figure}
\includegraphics[width=3.5in]{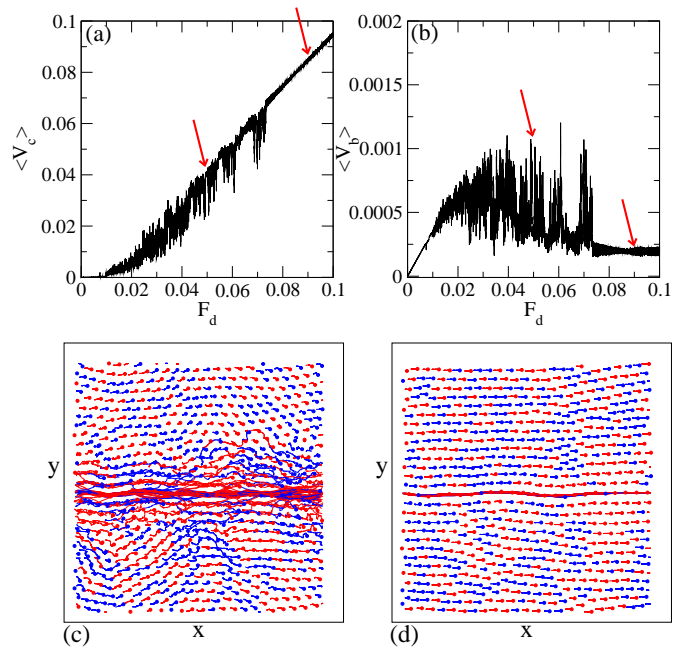}
\caption{ 
(a) $\langle V_{c}\rangle$ vs $F_{d}$ 
for a bidisperse system with $q_{1}/q_{2} = 1.6$ and $q_2=1.0$. 
At low drives, the system is crystalline and moves as a solid
unit. 
At decoupling, the system
enters a disordered regime, 
while at higher drives $F_d>0.075$,
there is a transition to a state with strongly localized shear and
reduced velocity fluctuations.
(b) $\langle V_{b}\rangle$ vs $F_{d}$ for the same system. 
The bulk velocity drops
almost to zero at the onset of the shear localization transition.    
(c,d) Particle positions (filled circles) and trajectories (lines) in the
same system for (c) the disordered decoupled state at
$F_d=0.05$ and (d) the strongly localized shear state at $F_d=0.09$.
Arrows in (a) and (b) indicate the drives at which the images in (c) and
(d) were obtained.
}
\label{fig:10}
\end{figure}

\subsection{Dynamic phases}
For bidisperse samples with large $q_1/q_2$,
we find an additional dynamical phase at high drives
where the shear band region 
becomes strongly localized again and the bulk 
particles lock together, similar to the behavior
found in the ordered regime.
In Fig.~\ref{fig:10}(a) we plot 
$\langle V_{c}\rangle$ versus $F_{d}$  
for a bidisperse sample with $q_{1}/q_{2} = 1.6$. 
For low $F_d$, the system is crystalline and strongly coupled, so that
all the particles move together.
As $F_d$ increases, the system becomes disordered 
and produces a large shear band
(illustrated in Fig.~\ref{fig:10}(c) 
for $F_{d} = 0.05$) associated with large velocity fluctuations.
At higher drives, however,
there is a transition to a strongly shear localized state
where only a single line of particles are moving while
the bulk particles lock together and moving at a very slow velocity, 
as shown in Fig.~\ref{fig:10}(d) for $F_{d} = 0.09$. 
The onset of this phase is accompanied by a drop
in the velocity fluctuations in Fig.~\ref{fig:10}(a).
In Fig.~\ref{fig:10}(b) we plot 
$\langle V_{b}\rangle$ versus $F_d$ for the same system,
showing that at the shear localization transition, the bulk velocity
drops to a value slightly above zero.
At the transition, $\langle V_{c}\rangle$ increases since the 
drag from the bulk particles on the particles in the driven channel 
is reduced.  
The shear localization occurs when the particles in the driven line
are moving sufficiently rapidly that they can no longer couple effectively
to the bulk particles.
This transition is a general feature that occurs in the disordered state.
Since the transition is dominated by fluctuations, 
the drive at which it occurs can vary significantly 
from one sample realization to another; however, 
on average the transition occurs 
at higher values of $F_{d}$ for increasing $q_{1}/q_{2}$ or decreasing $a$. 
The shear localized state has small velocity fluctuations with low noise that
is white ($\alpha=0$).

\begin{figure}
\includegraphics[width=3.5in]{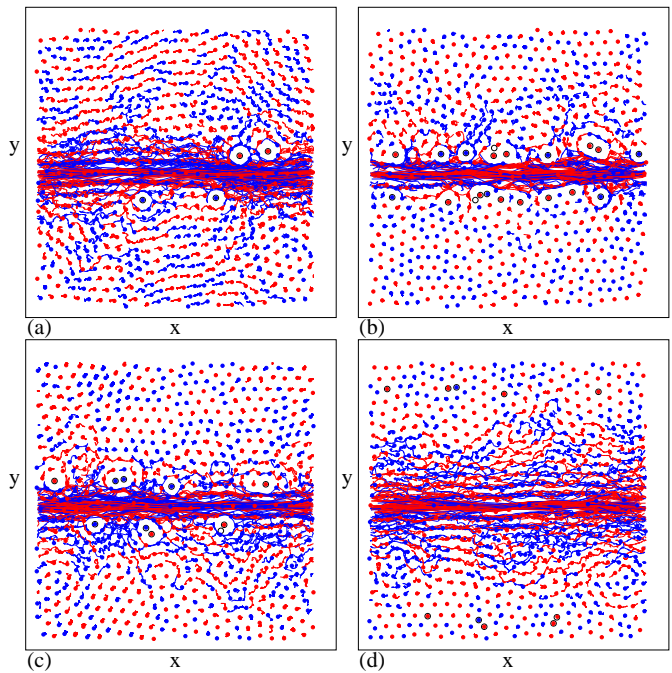}
\caption{ 
The particle positions (filled circles) and trajectories (lines) at 
a constant drive of $F_{d} = 0.05$ for 
a bidisperse system at $q_{1}/q_{2} = 1.8$ and $q_2=1.0$
with quenched disorder added in the form of localized pinning 
sites placed at an average distance
of $d_p$ from the driven line. 
The open circles are the locations of the pinning sites that each
capture one colloid. (a) $N_{p} = 4$ and $d_p=3a$.
(b) $N_{p} = 20$ and $d_p=3a$. As the number 
of pinning sites increases, the flow becomes more localized. 
(c) $N_p=10$ and $d_p=2a$. (d) $N_p=10$ and $d_p=18a$.
}
\label{fig:11}
\end{figure}

\section{Quenched Disorder and Shear Localization} 

We next examine the effects of adding quenched disorder or pinning 
sites in the bulk region.
In Fig.~11 we plot the particle trajectories in the disordered regime for
a bidisperse sample with $q_1/q_2=1.8$ at $F_d=0.05$.
The pinning sites are placed an average distance $d_p$ from the 
driven region.
Each pin captures a single colloid and the pinning force is sufficiently
strong that the particles do not depin over the range of $F_d$ we consider.
The repulsive nature
of the particle-particle interactions prevents unpinned particles from
closely approaching  pinned particles, leading to a reduction in the
density of particle trajectories in the vicinity of each pinned particle.
For a small number of pinning sites $N_p=4$, shown in Fig.~\ref{fig:11}(a)
for a sample with $d_p=3a$, 
the overall particle trajectories do not differ significantly from the
pin-free system; however, as the number of pinning sites increases,
the mixing region surrounding the driven line is reduced in width,
as illustrated in Fig.~\ref{fig:11}(b) for a sample with $N_p=20$ and $d_p=3a$.
Here, there is no longer any net motion of the bulk particles, indicating
a complete screening of the shear banding effect by the pinning.
This result shows that quenched disorder can 
produce strong shear localization.

\begin{figure}
\includegraphics[width=3.5in]{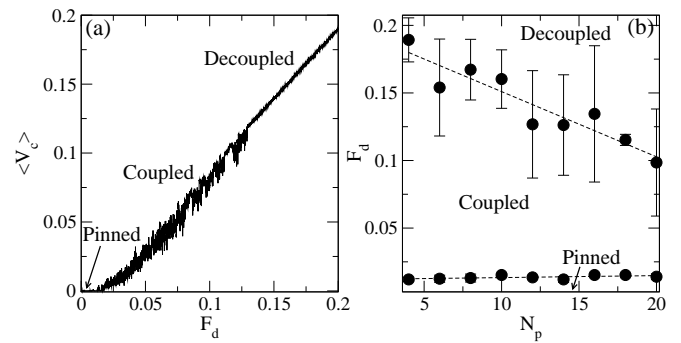}
\caption{ 
(a) $\langle V_{c}\rangle$ vs $F_{d}$ for a bidisperse system
at $q_1/q_2=1.8$ and $q_2=1.0$
with quenched disorder where $N_p=16$ and $d_p=2a$.
The pinned, coupled disordered, and high drive decoupled regimes
are labeled.
(b) Dynamic phase diagram for $F_d$ vs $N_p$ in bidisperse samples with
$q_{1}/q_{2} = 1.8$.
The pinned, coupled disordered, and high drive decoupled regimes are labeled.
As $N_p$ increases, the transition between the coupled and
decoupled phases drops to lower values of $F_d$.
}
\label{fig:12}
\end{figure}

In Fig.~\ref{fig:12}(a) we plot $\langle V_c\rangle$
versus $F_{d}$ for a bidisperse pinned system with $q_1/q_2=1.8$,
$N_p=16$, and $d_p=2a$ where we observe a
pinned regime, a coupled disordered flow regime, and a decoupled regime for
high drive similar to that found for the pin-free system in Fig.~\ref{fig:10}.
In the absence of pinning, at low drives the particles all move as a locked
solid; however, when pinning is added to the system, 
there is a true pinned phase 
at low drive where particle motion does not occur. In
Fig.~\ref{fig:12}(b) we plot the dynamic phase diagram 
for $F_{d}$ versus the number of pinning sites $N_{p}$. 
The transitions between the different phases show strong 
fluctuations from one realization
to another, and in Fig.~\ref{fig:12}(b) we show the transition line averaged over
several different disorder realizations.
The transition between the coupled and decoupled regimes drops to lower
$F_d$ with increasing $N_p$.
For a fixed drive, we find the interesting effect that the onset of the
decoupled regime is correlated with an increase in the velocity of the 
particles in the driven line.
Thus, by increasing the number of pinning sites, it is possible to cause the
driven particles to move at a higher velocity.
This effect arises due to the effective screening of the driven particles from
the bulk particles by the pinned particles.
A similar effect was observed for driving a single particle 
through a background of other particles
and pinning sites, where for high drives, a decoupling transition between
the driven and background particles leads to an increase in the velocity
of the driven particle.
In Fig.~\ref{fig:11}(c,d) we show the particle positions and trajectories for
samples where the number of pinning sites is fixed at $N_p=10$ for different
pin spacings of $d_p=2a$ [Fig.~\ref{fig:11}(c)] and $d_p=18a$ 
[Fig.~\ref{fig:11}(d)].
Here the motion is more localized for the smaller value of $d_p$.

\begin{figure}
\includegraphics[width=3.5in]{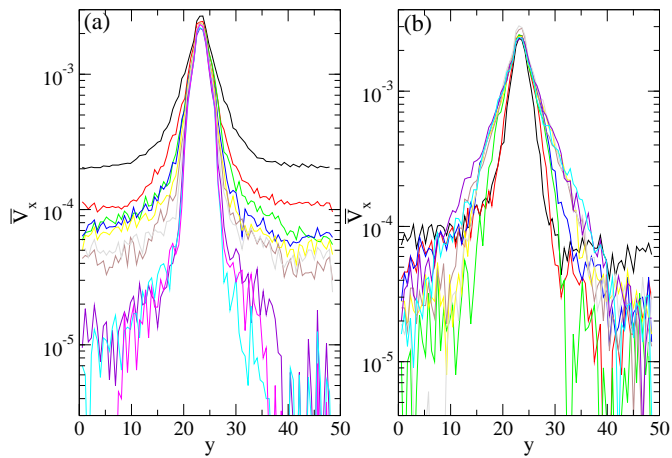}
\caption{ 
Velocity profiles $\bar V_x$ vs $y$ for bidisperse samples with
$q_1/q_2=1.8$ and $q_2=1.0$
at $F_d=0.024$.
(a) $d_p=2a$  and
$N_{p} = 0$, 4, 6, 8, 10, 12, 14, 16, 18, and $20$, 
from top center to bottom center.
Here $\bar V_x$ drops with increasing $N_p$ at all length scales.
(b) $N_{p} = 10$ and
$d_p=2a$, $4a$, $6a$, $8a$, $10a$, $12a$, $14a$, $16a$, and $18a$,
from bottom center to top center.
For the smallest $d_p$, the shear band is strongly localized.  
}
\label{fig:13}
\end{figure}
   
In Fig.~\ref{fig:13}(a) we show $\bar V_x(y)$ for bidisperse samples with $q_1/q_2=1.8$,
$d_p=2a$, and $F_d=0.024$ for different numbers of pinning sites ranging from
$N_p=0$ to $N_p=20$.
For the pin-free system with $N_p=0$, $\bar V_x$ far from the driven line
reaches a finite value since all of the particles in the system move.
As the number of pinning sites increases, 
$\bar V_x$ decreases at all length scales, 
with $\bar V_x$ in the driven line region decreasing more slowly than
$\bar V_x$ in the region far from the driven line.
For a sufficiently large number of pinning sites, 
$\bar V_x$ drops to zero for distances of $6a$ or greater away from
the driven line, indicating a complete screening of the shear band by
the pinned particles.
In Fig.~\ref{fig:13}(b) we show $\bar V_x$ for the same system with fixed $N_p=10$ and
varied $d_p$.
For increasing $d_p$, $\bar V_x$ in the bulk
gradually increases. 

\begin{figure}
\includegraphics[width=3.5in]{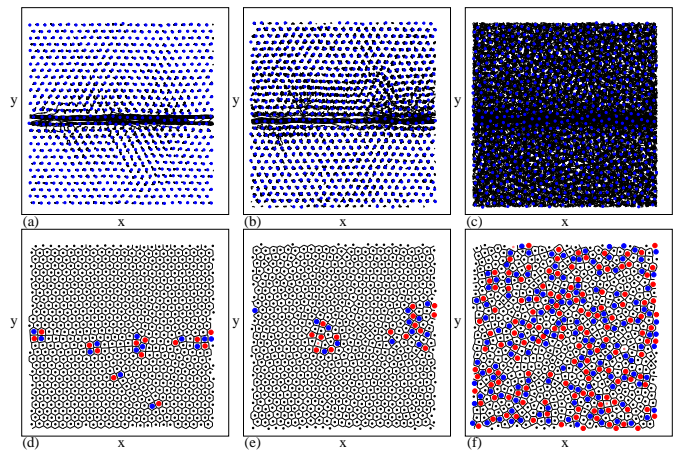}
\caption{ 
A monodisperse pin-free system with $q_1/q_2=1.0$, 
$q_2=1.0$, and $F_d=0.05$
at different temperatures.
(a,b,c) Dots: Particle positions.  
Lines: the
trajectories followed by the particles over a fixed time interval.
(d,e,f) Voronoi construction.  White polygons correspond to particles
with six neighbors, dark blue polygons to particles with five neighbors, and
light red polygons to particles with seven neighbors.
(a,d) $T=0.6$.  The dislocations are primarily located along the driven
line with their Burgers vectors aligned in the same direction.
(b,e) $T = 1.2$. 
The dislocations have begun to migrate into the bulk.
(c,f) $T = 1.8$, above the bulk melting temperature. 
The flow is disordered and topological defects have proliferated.
}
\label{fig:14}
\end{figure}

\section{Thermal Effects}
To study thermal effects, we
focus on the monodisperse system with $q_{1}/q_{2} = 1.0$
under fixed drive $F_d=0.05$, which
forms a crystalline state.
In Fig.~\ref{fig:14} we plot the particle positions and trajectories as well as
Voronoi constructions of snapshots of the particle positions
for increasing $T$. 
For low temperatures $T < 0.6$, 
the system remains ordered and the particle trajectories are  
one-dimensional. 
At $T = 0.6$, Fig.~\ref{fig:14}(a,d) shows that the trajectories near the 
driven line begin to mix, producing 5-7 paired defects, while the bulk
particles remain ordered.
At $T = 1.2$ in Fig.~\ref{fig:14}(b,e), the number of 5-7 defect pairs has
increased and some pairs begin to migrate from the driven line into the
bulk region, although the bulk remains mostly ordered.
For $T = 1.56$ the bulk begins to disorder and 
a widened shear band appears. Figure \ref{fig:14}(c,f) shows that at
$T = 1.8$ the bulk is strongly disordered. 

\begin{figure}
\includegraphics[width=3.5in]{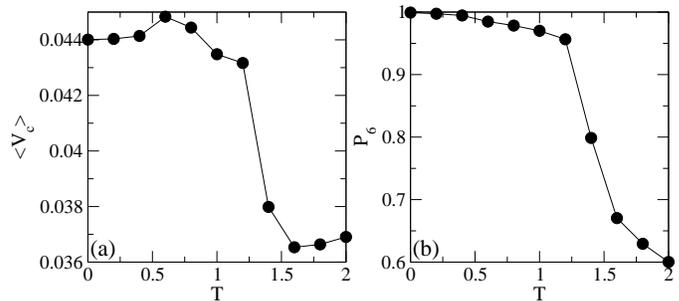}
\caption{ 
(a) $\langle V_{c}\rangle$ vs $T$ for a pin-free monodisperse 
system with $q_{1}/q_{2} = 1.0$ and $q_2=1.0$ at $F_{d} = 0.05$.  
(b) The fraction of six-fold coordinated particles $P_{6}$ vs $T$. 
The system melts near $T = 1.5$ and the
velocity of the particles in the driven line
drops at temperatures just below the bulk melting transition.  
}
\label{fig:15}
\end{figure}

In Fig.~\ref{fig:15}(a), the plot of $\langle V_{c}\rangle$ 
versus $T$ shows that the velocity drops
when the dislocations begin to migrate out from the driven line region.
There is a minimum in $\langle V_c\rangle$ at the bulk melting temperature.
Figure~\ref{fig:15}(b) shows the corresponding fraction of six-fold 
coordinated particles, $P_{6}=\sum_{i=1}^{N_c}\delta(z_i-6)$,
where $z_i$ is the coordination number of particle $i$.
For a perfect triangular lattice, $P_{6} = 1.0$.
We find that the drop in $P_{6}$, indicating bulk melting, 
occurs at a higher temperature than the drop in $\langle V_{c}\rangle$. 
In previous 
numerical studies of single driven probe 
particles moving through a crystalline background, 
it was also found that the
velocity drops when local dislocations 
near the driven probe particle occur, which typically begins
for temperatures below the bulk melting temperature \cite{14}. 
At much higher temperatures, the velocities gradually increase
again. This result shows that the driven line geometry 
can also be used to examine thermal melting properties.      

\section{Summary}

We study the effects of driving a quasi-one-dimensional region of 
particles through a two-dimensional system of Yukawa interacting particles.
In a monodisperse system a crystalline state forms, and as a function
of increasing external drive there
is a well defined transition from elastic flow, 
where all the particles move together, to a decoupled flow,
where the particles in the driven region decouple from the 
bulk particles and the bulk particles remain in a locked crystalline state.
The properties of this
elastic to decoupled transition are similar to those found
in studies of layered systems where only one layer is driven
and the other layer is dragged,  
such as a transformer geometry for bilayer superconductors 
or bilayer Wigner crystals. 
For bidisperse systems,
the bulk becomes disordered, 
and the driven line produces a local shear band 
and velocity gradient. 
The bulk particles near the driven line can be dragged with the driven
particles,
and the average velocity of the particles in the driven line 
decreases as the width of the shear band region increases. 
In the disordered regime, for increasing
drive we identify another decoupling transition where the bulk particles 
become locked, producing an elastic disordered solid, coinciding with the
shear band becoming very sharp. 
We show how the 
decoupling force, noise fluctuations, and average velocities 
can be correlated with bulk disordering transitions.  
We also consider the effects of adding quenched disorder or pinning, 
and find that the shear band region
becomes increasingly localized for increasing pinning density. 
This results in an interesting effect where the velocity of the driven particles
can be increased by adding more pinning sites to the system.
Such systems could be realized experimentally using
colloids or dusty plasmas 
with an optical drive applied along a one-dimensional channel, such as 
a laser applied along the edge of the sample. Variations
of this geometry could also be constructed in superconducting vortex
systems by embedding a single weak pinning channel 
in a bulk sample.   

\acknowledgments
This work was carried out under the auspices of the 
NNSA of the 
U.S. DoE
at 
LANL
under Contract No.
DE-AC52-06NA25396.
The work of A.~Lib{\' a}l was supported by a grant of the Romanian
National Authority for Scientific Research,
CNCS-UEFISCDI, project number PN-II-RU-TE-2011-3-0114.

\end{document}